%% file: main.tex
\newif\ifdraft
\newif\iffull
\newif\ifcomment
\newif\iflatexdiff
\newif\ifbibtex
\newif\ifpreprint
\preprinttrue    
\fulltrue        
\def\dvers{v0.4}
\def\snntitle{$\snn$}
\ifpreprint
\def\snntitle{$\snnbf$}
\fi
\def\dtitle{Long-range angular correlations of $\pi$, K and p\\in p--Pb collisions at \snntitle\ = \unit[5.02]{TeV}}
\def\stitle{Angular correlations of $\pi$, K and p in p--Pb collisions}
\ifpreprint
\documentclass[ALICE,manyauthors,12pt]{cernphprep}
\usepackage[comma,square,numbers,sort&compress]{natbib}
\usepackage{hyperref}
\else
\documentclass[final,3p,12pt]{elsarticle}
\biboptions{comma,square,numbers,sort&compress}
\usepackage{hyperref}
\fi
\usepackage{graphicx}  
\usepackage{dcolumn}   
\usepackage{bm}        
\usepackage{amssymb}   
\usepackage{amsfonts}
\usepackage{graphics}
\usepackage{grffile}   
\usepackage{epsfig}
\usepackage{units}
\usepackage[usenames]{color}
\usepackage[normalem]{ulem} 
\usepackage{color}
\usepackage[utf8]{inputenc}
\usepackage[T1]{fontenc}
\iflatexdiff
\RequirePackage{color}\definecolor{RED}{rgb}{1,0,0}\definecolor{BLUE}{rgb}{0,0,1}

\providecommand{\xout}[1]{{\protect\color{red}\sout{#1}}}

\fi
\ifpreprint
\else
 
\fi
\input{commands.tex}
\graphicspath{{./img/}}
\ifdraft
\usepackage{lineno}
\linenumbers
\modulolinenumbers[5]
\usepackage{fancyhdr}
\else
\renewcommand{\warn}[1]{}

\fi
\begin{document}
\newlength{\figlen}
\setlength{\figlen}{\linewidth}
\ifpreprint
\setlength{\figlen}{0.95\linewidth}
\begin{titlepage}
\PHnumber{2013-0115}                   
\PHdate{27 Jun 2013}                  
\title{\dtitle}
\ShortTitle{\stitle}
\Collaboration{ALICE Collaboration%
         \thanks{See Appendix~\ref{app:collab} for the list of collaboration members}}
\ShortAuthor{ALICE Collaboration} 
\ifdraft
\begin{center}
\today\\ \color{red}DRAFT \dvers\ \hspace{0.3cm} \$Revision: 150 $\color{white}:$\$\color{black}\vspace{0.3cm}
\end{center}
\fi
\else
\begin{frontmatter}
\title{\dtitle}
\iffull
\input{Alice_Authorlist_2013-Jun-25-ELSEVIER.tex}      
\else
\ifdraft
\author{ALICE Collaboration \\ \vspace{0.3cm}
\today\\ \color{red}DRAFT \dvers\ \hspace{0.3cm} \$Revision: 150 $\color{white}:$\$\color{black}}
\else
\author{ALICE Collaboration}
\fi
\fi
\fi
\begin{abstract}
Angular correlations between unidentified charged trigger particles and various species of charged associated particles (unidentified particles, pions, kaons, protons and antiprotons) are measured
by the ALICE detector in \pPb\ collisions at a nucleon--nucleon centre-of-mass energy of \unit[5.02]{TeV}
in the transverse-momentum range $0.3 < \pt < 4$ GeV/$c$.
The correlations expressed as associated yield per trigger particle are obtained in the pseudorapidity range $|\eta_{\rm lab}|<0.8$.
Fourier coefficients are extracted from the long-range correlations projected
onto the azimuthal angle difference and studied as a function of $\pt$ and in intervals of event
multiplicity. In high-multiplicity events, the second-order coefficient for protons, $v_2^{\rm p}$, is
observed to be smaller than that for pions, $v_2^\pi$, up to about $\pt=$~\unit[2]{\gevc}. To reduce correlations due to jets, the per-trigger yield measured in low-multiplicity events is subtracted from that in high-multiplicity events.
A two-ridge structure is obtained for all particle species. The Fourier decomposition of this structure shows that the second-order coefficients for pions and kaons are similar. The $v_2^{\rm p}$ is found to be smaller at low $\pt$ and larger at higher $\pt$ than $v_2^\pi$, with a crossing occurring at about \unit[2]{\gevc}. This is qualitatively similar to the elliptic-flow pattern observed in heavy-ion
collisions. A mass ordering effect at low transverse momenta is consistent with expectations from hydrodynamic model calculations assuming a collectively
expanding system.
\ifdraft
\ifpreprint
\end{abstract}
\end{titlepage}
\else
\end{abstract}
\end{frontmatter}
\newpage
\fi
\fi
\ifdraft
\thispagestyle{fancyplain}
\else
\end{abstract}
\ifpreprint
\end{titlepage}
\else
\end{frontmatter}
\fi
\fi
\setcounter{page}{2}

\input{content.tex}

\ifpreprint
\iffull
\newenvironment{acknowledgement}{\relax}{\relax}
\begin{acknowledgement}
\section*{Acknowledgements}
\input{acknowledgements_march2013.tex}        
\end{acknowledgement}
\ifbibtex
\bibliographystyle{utphys}
\bibliography{biblio}{}
\else
\input{refpreprint.tex}
\fi
\newpage
\appendix
\section{The ALICE Collaboration}
\label{app:collab}
\input{Alice_Authorlist_2013-Jun-25.tex}      
\else
\ifbibtex
\bibliographystyle{utphys}
\bibliography{biblio}{}
\else
\input{refpreprint.tex}
\fi
\fi
\else
\iffull
\vspace{0.5cm}
\input{acknowledgements_march2013.tex}
\input{refpaper.tex}
\else
\ifbibtex
\bibliographystyle{utphys}
\bibliography{biblio}{}
\else
\input{refpaper.tex}
\fi
\fi
\fi
\end{document}

%% file: commands.tex
\newcommand{\gevc}         {GeV/\ensuremath{c}}
\newcommand{\gevcc}         {GeV/\ensuremath{c^2}}

\newcommand{\ITS}          {\rm{ITS}}
\newcommand{\TOF}          {\rm{TOF}}
\newcommand{\ZNA}          {\rm{ZNA}}
\newcommand{\ZNC}          {\rm{ZNC}}
\newcommand{\ZDC}          {\rm{ZDC}}
\newcommand{\ZDCs}         {\rm{ZDCs}}

\newcommand{\TPC}          {\rm{TPC}}
\newcommand{\VZERO}        {\rm{VZERO}}
\newcommand{\VZEROA}       {\rm{VZERO-A}}
\newcommand{\VZEROC}       {\rm{VZERO-C}}
\newcommand{\TZERO}        {\rm{T0}}

\newcommand{\dedx}         {d$E$/d$x$}

\newcommand{\pp}           {pp}

\newcommand{\Aa}           {\mbox{A--A}}
\newcommand{\PbPb}         {\mbox{Pb--Pb}}

\newcommand{\pPb}          {\mbox{p--Pb}}

\newcommand{\dNdeta}       {\mathrm{d}N_\mathrm{ch}/\mathrm{d}\eta}

\newcommand{\s}            {\ensuremath{\sqrt{s}}}
\newcommand{\pt}           {\ensuremath{p_{\mathrm{T}}}{ }}

\newcommand{\snn}          {\ensuremath{\sqrt{s_{\mathrm{NN}}}}}
\newcommand{\snnbf}        {\ensuremath{\mathbf{{\sqrt{s_{\mathbf NN}}}}}}

\newcommand{\avg}[1]       {\ensuremath{\left\langle#1\right\rangle}}

\newcommand{\dd}           {\ensuremath{\mathrm{d}}}

\newcommand{\Dphi}         {\ensuremath{\Delta\varphi}}
\newcommand{\Deta}         {\ensuremath{\Delta\eta}}
\newcommand{\Ntrig}        {\ensuremath{N_{\mathrm{trig}}}}
\newcommand{\Nassoc}       {\ensuremath{N_{\mathrm{assoc}}}}
\newcommand{\dNassoc}      {\ensuremath{\frac{\dd^2N_{\mathrm{assoc}}}{\dd\Deta\dd\Dphi}}}

\newcommand{\Fig}[1]       {Fig.~\ref{#1}}

\newcommand{\Sect}[1]      {Sect.~\ref{#1}}

\newcommand{\Eq}[1]        {Eq.~\ref{#1}}

\newcommand{\Ref}[1]       {Ref.~\citenum{#1}}

\newcommand{\red}[1]       {\textcolor{red}{#1}}

\newcommand{\warn}[1]      {{\small\textbf{\red{(!}\footnote{\textbf{\red{(!)}}~#1}\red{)}}}\marginpar{\textbf{\red{---}}}}

\newcommand{\com}[1]       {}

\iflatexdiff

\renewcommand{\xout}[1]    {\textcolor{red}{\sout{#1}}}
 
\else

\renewcommand{\xout}[1]    {}
\fi

%% file: content.tex

\section{Introduction}
\label{sec:intro}
Measurements of the correlations of two or more particles are a powerful tool to study the
underlying mechanism of particle production in collisions of hadrons and nuclei at high energy.
These studies often involve measuring the distributions of relative angles $\Dphi$ and $\Deta$, where $\Dphi$ and $\Deta$ are the differences in azimuthal angle~$\varphi$
and pseudorapidity~$\eta$ between the directions of two particles.

In minimum-bias proton--proton~(\pp) collisions, the correlation at ($\Dphi \approx 0$, $\Deta \approx 0$) is dominated
by the ``near-side'' jet peak, and at $\Dphi \approx \pi$ by the recoil or ``away-side'' structure due to particles originating
from jet fragmentation~\cite{Wang:1992db}.  In nucleus--nucleus~(\Aa) collisions additional structures along the $\Deta$ axis
emerge on the near and away side in addition to the jet-related
correlations~\cite{Adams:2004pa, Alver:2008gk, Alver:2009id, Abelev:2009qa, Chatrchyan:2011eka, Aamodt:2011by,
  Agakishiev:2011pe, Chatrchyan:2012wg, ATLAS:2012at, Aamodt:2011vk, Adare:2011tg, Abelev:2012di, Chatrchyan:2012zb}.
These ridge-like structures persist over a long range in $\Deta$.
The shape of these $\Dphi$ correlations can be studied via a Fourier decomposition~\cite{Voloshin:1994mz}.
The second and third order terms are the dominant harmonic coefficients $v_{n}$~\cite{Aamodt:2011vk, Adare:2011tg, Aamodt:2011by, Chatrchyan:2011eka, Chatrchyan:2012wg, Abelev:2012di,
  Chatrchyan:2012zb, ATLAS:2012at}.  The $v_n$ coefficients can be related to the collision geometry and density
fluctuations of the colliding nuclei~\cite{Ollitrault:1992bk,Alver:2010gr} and to the transport properties of the created matter in hydrodynamic models~\cite{Alver:2010dn, Schenke:2010rr, Qiu:2011hf}.

In \pp\ collisions at a centre-of-mass energy $\s=$~\unit[7]{TeV} the emergence of similar long-range ($2<|\Deta|<4$) near-side ($\Dphi \approx 0$) correlations was reported in
events with significantly higher-than-average particle multiplicity~\cite{Khachatryan:2010gv}.
This was followed by the observation of the same structure in high-multiplicity
proton--lead~(\pPb) collisions at a nucleon--nucleon centre-of-mass energy $\snn=$~\unit[5.02]{TeV}~\cite{CMS:2012qk}.
Subsequent measurements in \pPb\ collisions employed a procedure for
removing the jet contribution by subtracting the correlations extracted from low-multiplicity events, revealing essentially the same long-range structures on the away side in high-multiplicity
events~\cite{alice_pa_ridge,atlasridge}.  Evidence of similar long-range structures in high-multiplicity deuteron--gold
collisions at $\snn=$~\unit[0.2]{TeV} has also been reported~\cite{Adare:2013piz}.  In all
cases~\cite{alice_pa_ridge,atlasridge,Adare:2013piz}, the transverse-momentum~($\pt$) dependence of the extracted $v_2$
and $v_3$ coefficients is found to be similar to that measured in \Aa\
collisions.  Recent measurements involving two- and four-particle correlations~\cite{Aad:2013fja,Chatrchyan:2013nka}
revealed that the $\pt$-integrated $v_3$ in \pPb\ collisions is the same as in \PbPb\ collisions at the same midrapidity
multiplicity. Further, genuine four-particle correlations utilizing cumulants~\cite{Bilandzic:2010jr} lead to non-zero $v_2$ coefficients that are somewhat smaller
than those extracted from two-particle correlations, and smaller than those in \PbPb\ collisions at the same
midrapidity multiplicity.

The ridge structures in high-multiplicity \pp\ and \pPb\ events have been attributed to mechanisms that
involve initial-state effects, such as gluon saturation and colour connections forming along the longitudinal
direction~\cite{Arbuzov:2011yr, Dusling:2012cg, Dusling:2012wy, Kovchegov:2012nd,Dusling:2013oia,Bzdak:2013zma} and final-state effects, such as parton-induced interactions~\cite{Wong:2011qr, Strikman:2011cx,
Alderweireldt:2012kt}, and collective effects arising in a high-density system possibly formed
in these collisions~\cite{Avsar:2010rf, Werner:2010ss, Deng:2011at, Avsar:2011fz, Bozek:2011if, Bozek:2012gr,
Bozek:2013uha, Shuryak:2013ke}.

A dense, highly interacting system exhibiting radial
collective (hydrodynamic) flow, as the one formed in central \Aa\ collisions,
leads to a characteristic particle-species dependent modification of the $\pt$ spectra of identified
particles as observed in~\cite{Abelev:2008ab, Adare:2013esx, alice_pa_spectra}.
Furthermore, the correlations of identified particles can be used to investigate the presence of a collective expansion since
the $v_2$ of lighter identified particles should be larger than that of heavier particles at the same $\pt$ \cite{Huovinen:2001cy}.
Indeed, in \Aa\ collisions, for $\pt<$~\unit[2]{\gevc}, $v_2$ exhibits a particle-mass dependence~\cite{Adler:2003kt, Adams:2004bi, Heinz:2011kt} as predicted by hydrodynamic model
calculations~\cite{Huovinen:2001cy, Shen:2011eg}.
At intermediate $\pt$ ($2<\pt<$~\unit[8]{\gevc}) the $v_2$ of mesons is smaller than that of
baryons even at similar particle mass~\cite{Abelev:2007qg, Adare:2012vq, Abelev:2012di}, which may be attributed to quark coalescence~\cite{PhysRevLett.91.092301, PhysRevLett.90.202302, PhysRevLett.90.202303}.

In this letter, measurements of the $v_2$ of pions, kaons and protons\footnote{Pions, kaons and protons, as well as the symbols $\pi$, K and p, refer to the sum of particles and antiparticles.} in \pPb\ collisions at $\snn=$ \unit[5.02]{TeV} are presented. These results are obtained from two-particle correlations and extend the characterization of the double ridge observed in \pPb\ collisions.

\section{Experimental setup}
\label{sec:setup}
Data from the 2013 \pPb\ run of the LHC for collisions of \unit[4]{TeV} protons and \unit[1.58]{TeV} per nucleon lead ions, resulting in a centre-of-mass energy of $\snn=$~\unit[5.02]{TeV}, are used in the presented analysis. The nucleon--nucleon centre-of-mass system is offset with respect to the ALICE laboratory system by $-0.465$ in rapidity, i.e. in the direction of the proton beam.
In the following, $\eta$ denotes the pseudorapidity in the laboratory system.

A detailed description of the ALICE detector can be found in \Ref{Aamodt:2008zz}.
The main subsystems used in the present analysis are the Inner Tracking System~(\ITS), the
Time Projection Chamber~(\TPC) and the Time Of Flight detector~(TOF). These have a common acceptance of
$|\eta| < 0.9$ and are operated inside a solenoidal magnetic field of \unit[0.5]{T}.
The \ITS\ consists of six layers of silicon detectors for vertex finding and tracking.
The \TPC\ is the main tracking detector and provides particle identification by measuring the specific energy loss \dedx.
The TOF and \TZERO\ detectors are used to identify particles by measuring their flight time.
The \TZERO\ detectors have a pseudorapidity coverage of $-3.3 < \eta < -3.0$ and $4.6 < \eta < 4.9$.
The VZERO detector, two arrays of 32 scintillator tiles each, covering $2.8<\eta<5.1$~(\VZEROA) and $-3.7<\eta<-1.7$~(\VZEROC), was used for triggering and
event selection. The trigger required a coincidence of signals in both \VZEROA\ and \VZEROC. In addition, two neutron Zero Degree Calorimeters~(\ZDCs) located at \unit[112.5]{m}~(\ZNA) and \unit[$-112.5$]{m}~(\ZNC) from the interaction point are used in the event selection.
The \VZEROA, which is located in the flight direction of the Pb ions, is used to define event classes corresponding to different
particle-multiplicity ranges. Alternatively, the energy deposited in the \ZNA\ (originating from neutrons from the Pb nucleus) is used in defining the event-multiplicity classes.
All these detector systems have full azimuthal coverage.

\section{Event, track selection and particle identification}
\label{sec:selection}
The event selection for this analysis is based on signal amplitudes and their arrival times measured with the \VZERO\ and \ZDC\ detectors.
It is identical to the selection described in \Ref{ALICE:2012xs} which accepts 99.2\% of all non-single-diffractive collisions.
The collision-vertex position is determined with tracks reconstructed in the \ITS\ and \TPC\ as described
in \Ref{Abelev:2012eq}. The vertex reconstruction algorithm is fully efficient for events with at least
one reconstructed primary track within $|\eta|<1.4$~\cite{perfpaper}.
The position of the reconstructed vertex along the beam
direction ($z_{\rm vtx}$) is required to be within \unit[10]{cm} of the detector centre.
About $10^8$ events, corresponding to an integrated luminosity of about \unit[50]{$\mu$b$^{-1}$}, pass these event selection criteria and are used for the analysis.

The analysis uses charged-particle tracks reconstructed in the \ITS\ and \TPC\ with $0.3<\pt<$~4 \gevc\ and in a fiducial region of $|\eta|<0.8$ to exclude non-uniformities at the detector edges. As a first step, track selection criteria on the number of space points and on
the quality of the track fit in the \TPC\ are applied \cite{ALICE:2012ab}. Tracks are additionally required to have at least
one hit in the two innermost layers of the \ITS\ and to have a Distance of Closest Approach (DCA) to the reconstructed collision vertex smaller than \unit[2]{cm} in the longitudinal direction.
In the transverse direction, a cut at $7\sigma_{\rm dca}$ is applied, where $\sigma_{\rm dca}$ is the $\pt$-dependent transverse impact-parameter resolution (\unit[30--200]{$\mu$m} from highest to lowest $\pt$ in the considered range) \cite{ALICE:2012ab}.
To study the effect of contamination by secondary particles, the transverse DCA cut is varied between 3 and $21\sigma_{\rm dca}$. For the scalar-product method analysis, discussed below, tracks without a hit in the two innermost layers of the ITS, but having a hit in the third layer, are retained, to achieve a more uniform $\varphi$ acceptance.
For tracks with \pt>~\unit[0.5]{\gevc} a signal in the TOF is required for particle identification.
The track selection is varied in the analysis as a consistency check.

Particle identification is performed using the specific energy loss \dedx\ in the \TPC\ and the time of flight measured with the \TOF\ (for $\pt >$~\unit[0.5]{\gevc}). A truncated mean procedure is used in order to reduce the Landau tail of the energy loss distribution in the \TPC\ (60\% of the measured clusters are kept) \cite{Abelev:2013vea}. The \dedx\ resolution is 5--6\%, depending upon the number of associated space points in the \TPC.
The resolution of the time of flight is given by the detector resolution and the resolution of the collision time measurement. The collision time can be computed utilizing three different methods: a) from the \TZERO\ detectors, b) from a combinatorial algorithm which uses the TOF measurement itself, or c) from the average collision time~\cite{Akindinov:2013tea} (only used in few low-multiplicity events where the first two measurements are missing). The corresponding time of flight resolution is about \unit[85]{ps} for high-multiplicity events and about \unit[120]{ps} for low-multiplicity events.

Based on the difference (expressed in units of the resolution $\sigma$) between the measured signal and the expected signal for $\pi$, K, or p in the \TPC\ and \TOF, a combined $N_{\sigma, \rm PID}^2 = N_{\sigma, \rm TPC}^2 + N_{\sigma, \rm TOF}^2$ is computed.
For a given species, particles are selected with a circular cut in the $N_{\sigma, \rm TPC}$ and $N_{\sigma, \rm TOF}$ space by $N_{\sigma, \rm PID} < 3$. In the region where the areas of two species overlap, the identity corresponding
to the smaller $N_{\sigma, \rm PID}$ is assigned. For $\pt$ less than \unit[0.5]{\gevc} only a few tracks have an associated signal in the \TOF\ and $N_{\sigma, \rm PID} = N_{\sigma, \rm TPC}$ is used.
This strategy provides track-by-track identification with high purity over the momentum region considered in this letter: $0.3 < \pt <$~\unit[4]{\gevc} for pions, $0.3 < \pt <$~\unit[3]{\gevc} for kaons and $0.5 < \pt <$~\unit[4]{\gevc} for protons.
To assess the systematic uncertainty related to the particle identification, the selection is changed to $N_{\sigma, \rm PID} < 2$. Furthermore, an exclusive identification is used, in which the tracks that are within the $N_{\sigma, \rm PID}$ overlap area are rejected.
Both selections reduce the misidentification rate.

The efficiency and purity of the primary charged-particle selection are estimated from a Monte Carlo (MC)
simulation using the DPMJET version 3.05 event generator~\cite{Roesler:2000he} with particle transport
through the detector using GEANT3~\cite{geant3ref2} version 3.21 which contains an improved description of the $\bar{\rm p}$ inelastic cross section \cite{alice_pa_spectra}. The cross sections for the interactions of negative kaons at low $\pt$ with the detector material are known to not be correctly reproduced in GEANT3~\cite{Abelev:2013vea}. Therefore, the efficiency extracted from GEANT3 was scaled with a factor computed with a dedicated FLUKA~\cite{Battistoni:2007zzb} simulation as discussed in \cite{Abelev:2013vea}. This correction ranges from about 10\% to about 1\% from the lowest to the highest \pt interval considered.
The efficiency and acceptance for track reconstruction depends on particle species and is
61--87\% for the $\pt$ range \unit[0.5--1]{\gevc}, and 79--86\% at $\pt=$~\unit[4]{\gevc}.
The additional efficiency factor for a track having an associated signal in the \TOF\ and being correctly identified is
about 59\%, 43\% and 48\% for the \pt range \unit[0.5--1]{\gevc} for $\pi$, K and p, respectively, and saturates at about 63\% at $\pt=$~\unit[2]{\gevc} for all the species.
These numbers are independent of the event multiplicity.

The remaining contamination from secondary particles due to interactions in the detector material and due to weak decays decreases from about 20\% to 1\% for protons in the $\pt$ range from 0.5 to \unit[4]{\gevc} and from about 4\% to 0.5\% for pions in the $\pt$ range from 0.3 to \unit[4]{\gevc} while it is negligible for kaons.
The contribution from fake tracks from random associations of detector signals is negligible.
The contamination from misidentified particles is significant only for
kaons above \unit[1.5]{\gevc} and is less than 15\%. Corrections for these effects are discussed in Section~\ref{sec:twopartfunc}.

The two-particle correlations are studied by dividing the selected
events into four classes defined as fractions of the analyzed event
sample, based on the charge deposition in the \VZEROA\ detector, and denoted
``0--20\%'', ``20--40\%'', ``40--60\%'', ``60--100\%'' from the highest to the lowest multiplicity.
The event-class definitions are shown in
Table~\ref{tab:multclasses} together with the corresponding mean charged-particle
multiplicity densities within $|\eta|<0.5$ ($\avg{\dNdeta}$).
The multiplicity estimate is corrected for detector acceptance,
track-reconstruction efficiency and contamination.
Contrary to our earlier measurement of $\avg{\dNdeta}$~\cite{ALICE:2012xs}, the value here is not corrected for trigger and vertex-reconstruction efficiency.
Also shown is the mean number of primary charged particles with $\pt>$~\unit[0.5]{\gevc} within the pseudorapidity range $|\eta|<0.8$ ($\avg{N_{\rm trk}}$). This is measured by applying the track selection described above and is corrected for the detector acceptance, track reconstruction efficiency and contamination.

\begin{table}[bht!f] \centering
  \caption{\label{tab:multclasses}
    Event classes defined as fractions of the analyzed event sample and
    their corresponding $\avg{\dNdeta}$ within $|\eta|<0.5$
    and the mean numbers of charged particles within $|\eta|<0.8$ and $\pt>$~\unit[0.5]{\gevc}. The uncertainties are only systematic as the statistical uncertainties are negligible.
    }
  \begin{tabular}{cccc}
    \hline
    Event        & \VZEROA & $\avg{\dNdeta}|_{|\eta|<0.5}$  & $\avg{N_{\rm trk}}|_{|\eta|<0.8}$ \\
    class        & range (a.u.)    & $\pt>$~\unit[0]{\gevc} & $\pt>$~\unit[0.5]{\gevc} \\
    \hline
      60--100\% & $<52$      & $\rm 7.1 \pm 0.2\ $  & $\rm 4.6  \pm 0.2$ \\
      40--60\%  & $52-89$    & $\rm 16.1 \pm 0.4\ $  & $\rm 11.5 \pm 0.4 $ \\
      20--40\%  & $89-142$   &$\rm 23.2  \pm 0.5\ $ & $\rm 17.3 \pm 0.6 $ \\
       0--20\%  & $>142$     & $\rm  35.6 \pm 0.8\ $ & $\rm 27.5 \pm 1.0 $ \\
    \hline
  \end{tabular}
\end{table}

\section{Analysis}
\label{sec:twopartfunc}

The $v_n$ coefficients are extracted using two methods, referred to in the following as two-particle correlations and scalar product. In two-particle correlations both particles are taken from the same $\pt$ interval, while in the scalar-product method, particles from a certain \pt interval are correlated with particles from the full \pt range.
Comparing the results of these two methods can address to what extent the Fourier coefficients of two-particle correlations factorize into the product of the coefficients of the corresponding single-particle angular distributions.
In particular, these results should agree if the measurement is dominated by correlations of each of the particles with a common plane.

\subsection{Two-particle correlations}
The correlation between two particles (denoted trigger and associated particle) is measured
as a function of the azimuthal angle difference $\Dphi$ (defined within $-\pi/2$ and $3\pi/2$) and pseudorapidity
difference $\Deta$~\cite{alice_pa_ridge}. While the trigger particles are in all cases unidentified charged particles, the analysis is done separately for unidentified charged associated particles (denoted $h-h$)
and for associated charged pions, kaons and protons (denoted $h-\pi$, $h-{\rm K}$ and $h-{\rm p}$, respectively). The correlation is
expressed
in terms of the associated yield per trigger particle where both particles are from the same $\pt$ interval in a fiducial region of $|\eta|<0.8$:
\begin{equation}
\frac{1}{\Ntrig} \dNassoc = \frac{S(\Deta,\Dphi)}{B(\Deta,\Dphi)} \label{pertriggeryield}
\end{equation}
where $\Ntrig$ is the total number of trigger particles in the event class and $\pt$ interval.
The signal distribution
$S(\Deta,\Dphi) = 1/\Ntrig\ \dd^2N_{\rm same}/\dd\Deta\dd\Dphi$
is the associated yield per trigger particle for particle pairs from the same event.
The background distribution $B(\Deta,\Dphi) = \alpha\ \dd^2N_{\rm mixed}/\dd\Deta\dd\Dphi$
corrects for pair acceptance and pair efficiency.
It is constructed by correlating the trigger particles in one event with the associated particles from
other events in the same event class and within the same \unit[2]{cm}-wide $z_{\rm vtx}$ interval
(each event is mixed with about 5--20 events).
The background distribution is normalized with a factor $\alpha$ which is chosen such that it is unity for pairs where
both particles travel in approximately the same direction (i.e.\ $\Dphi\approx 0,\ \Deta\approx 0$).
The yield defined by \Eq{pertriggeryield} is constructed for each $z_{\rm vtx}$ interval to account for differences in pair acceptance and in pair efficiency as a function of $z_{\rm vtx}$.
After efficiency correction, described below, the final per-trigger yield is obtained by calculating the average of the $z_{\rm vtx}$ intervals weighted by $\Ntrig$.

A selection on the opening angle of the particle pairs is applied to avoid a bias due to the reduced efficiency for pairs with small opening angles. Pairs are required to have a separation of $|\Delta\varphi^{*}_{\rm min}|>$~\unit[0.02]{rad} or $|\Deta|>0.02$, where $\Delta\varphi^*_{\rm min}$  is
the minimal azimuthal distance at the same radius between the two tracks within the active
detector volume after accounting for the bending in the magnetic field.
Furthermore, correlations induced by secondary particles from neutral-particle decays are suppressed by cutting on the invariant mass ($m_{\rm inv}$) of the particle pair. Pairs are removed which are likely to stem from a $\gamma$-conversion ($m_{\rm inv} <$~\unit[0.04]{\gevcc}), or a K$^0_s$ decay ($|m_{\rm inv} - m({\rm K}^0)| <$~\unit[0.02]{\gevcc}) or a
$\Lambda$ decay ($|m_{\rm inv} - m(\Lambda)| <$~\unit[0.02]{\gevcc}).
The contribution from decays where only one of the decay products has been reconstructed is estimated by varying the DCA cut as discussed above and found to be only relevant for protons (due to $\Lambda$ feed-down) below \unit[2]{\gevc}.
Weak decays of heavier particles give a negligible contribution.

Each trigger and each associated particle is weighted with a correction factor that accounts for detector acceptance, reconstruction efficiency and contamination by secondary particles. For the identified associated particles this correction factor also includes the particle-identification efficiency. These corrections are applied as a function of $\eta$, $\pt$ and $z_{\rm vtx}$. The $v_n$ coefficients extracted below are expected to be insensitive to single-particle weights as they are relative quantities. Indeed, the coefficients with and without these corrections are identical.

The effect of wrongly identified particles, particularly relevant for pions misidentified as kaons, is corrected
by subtracting the measured $h-\pi$ per-trigger yield from the measured $h-{\rm K}$ per-trigger yield scaled with the
misidentification fraction (percentage of pions identified as kaons) extracted from MC. Similarly, this is done for the contamination of the $h-{\rm p}$ per-trigger yield.
The correction procedure is validated by applying it on simulated events and comparing the $v_n$ coefficients with the input MC.

Compared to our previous publication~\cite{alice_pa_ridge}, the following analysis details have changed: a) event-multiplicity classes are defined with the \VZEROA\ instead of the combination of both \VZERO\ detectors because the \VZEROA\ is in the direction of the Pb beam and is thus more sensitive to the fragmentation of the Pb nucleus; b) the fiducial volume is reduced to $|\eta| < 0.8$ due to the use of particle identification; and c) the
condition that the associated transverse momentum has to be smaller than the trigger transverse momentum is not applied.

Fourier coefficients can be extracted from the $\Dphi$ projection of the per-trigger yield by
a fit with:
\begin{equation}
  \frac{1}{\Ntrig} \frac{\dd \Nassoc}{\dd\Dphi} = a_0 + 2\,a_1 \cos \Dphi + 2\,a_2 \cos 2\Dphi + 2\,a_3 \cos 3\Dphi.
  \label{fitfunction1}
\end{equation}
From the relative modulations $V_{n\Delta}^{h-i}\{{\rm 2PC}\} = a_n^{h-i} / a_0^{h-i}$, where $a_{n}^{h-i}$ is the $a_{n}$ extracted from $h-i$ correlations, the $v_{n}^{i}\{{\rm 2PC}\}$ coefficient of order $n$ for a particle species $i$ (out of $h$, $\pi$, K, p) are then defined as:
\begin{eqnarray}
  v_{n}^{h}\{{\rm 2PC}\} = \sqrt{V_{n\Delta}^{h-h}} \hspace{2cm} v_{n}^{i}\{{\rm 2PC}\} = V_{n\Delta}^{h-i} / \sqrt{V_{n\Delta}^{h-h}}. \label{vn}
\end{eqnarray}
In the case that each of the particles is correlated with a common plane,
the $v_{n}^{i}\{{\rm 2PC}\}$ are the Fourier coefficients of the corresponding single-particle angular distributions.

\subsection{Scalar-product method}

Alternatively, the scalar-product method~\cite{Voloshin:2008dg} is used to extract the $v_n$ coefficients:
\begin{equation}
v_{n}\{{\rm SP}\} = \frac{\langle \Re\left({\bf u}_{n, k}{\bf Q}_{n}^{*}\right)/M \rangle}{\sqrt{\langle \Re\left({\bf Q}_{n}^{a} {{\bf Q}_{n}^b}^*\right)/(M^{a}M^{b}) \rangle}},
\label{eq:mth_sp}
\end{equation}
where ${\bf u}_{n, k}=\exp{in\varphi_k}$ is the unit vector of the particle of interest $k$, ${\bf Q}_{n} = \sum_l \exp{in\varphi_l}$ is the event flow
vector, $M$ is the event multiplicity, and $n$ is the harmonic number. The full event is divided into
two independent sub-events $a$ and $b$ composed of tracks from different pseudorapidity intervals with flow vectors ${\bf Q}_{n}^{a}$
and ${\bf Q}_{n}^{b}$ and multiplicities $M^{a}$ and $M^{b}$.
The angle brackets denote an average over all particles
in all events, $\Re$ the real part of the scalar product and $^*$ the complex conjugate.

To determine ${\bf Q}_{n}$, either $h$, $\pi$, K or p are taken as particles of interest from a $\pt$ interval and correlated with all unidentified particles from the full $\pt$ range (reference particles).
The two sub-events $a$ and $b$ are defined within the pseudorapidity range $-0.8<\eta<-0.4$ and $0.4<\eta<0.8$, respectively. The particle of interest is taken from $a$ and the reference particles from $b$ and vice-versa.
This results in a pseudorapidity gap of $|\Deta|>0.8$ which reduces correlations from jets and resonance
decays.

Non-uniformities in the acceptance are corrected using the prescription in \cite{Bilandzic:2010jr}. This correction is less than 5\%.
As above, the coefficients can be shown to be insensitive to single-particle effects. The contamination by secondary particles from weak decays is estimated varying the DCA cut, as detailed above.
The influence of misidentified particles is corrected for, e.g., in the case of kaons by subtracting the $v_n^\pi$ from $v_n^{\rm K}$ taking the particle ratios from the data and the misidentification fraction extracted from MC into account.
The correction method is validated on simulated events.

Table~\ref{tab:syst} summarizes the uncertainties related to the $v_2$ measurements. Details of the separate contributions are given in the text where they are introduced.

\begin{table}[bht!f] \centering
  \caption{Summary of main systematic uncertainties. The uncertainties depend on $\pt$ and multiplicity class and vary within the given ranges. $v_2\{{\rm 2PC, sub}\}$ is introduced in \Sect{sec:results}. \label{tab:syst}
    }
  \begin{tabular}{lccc}
    \hline
    Source					& $v_2\{{\rm 2PC}\}$ & $v_2\{{\rm SP}\}$ & $v_2\{{\rm 2PC, sub}\}$ \\
    \hline
      Track selection and efficiencies 		& 2--20\% 	& 2--20\%	& 0--3\% \\
      Particle identification	 		& 2--6\% 	& 2--3\%	& 2--7\% \\
      Contamination by weak decays (only p)	& 0--10\% 	& 0--10\%	& 0--4\% \\
      Residual jet contribution			& ---		& ---		& 3--10\% \\
    \hline
      Sum			 		& 2--20\% 	& 	2--20\%	& 3--14\% \\
    \hline
  \end{tabular}
\end{table}

\section{Results}
\label{sec:results}

\begin{figure}[ht!f]
\centering
\includegraphics[width=\textwidth]{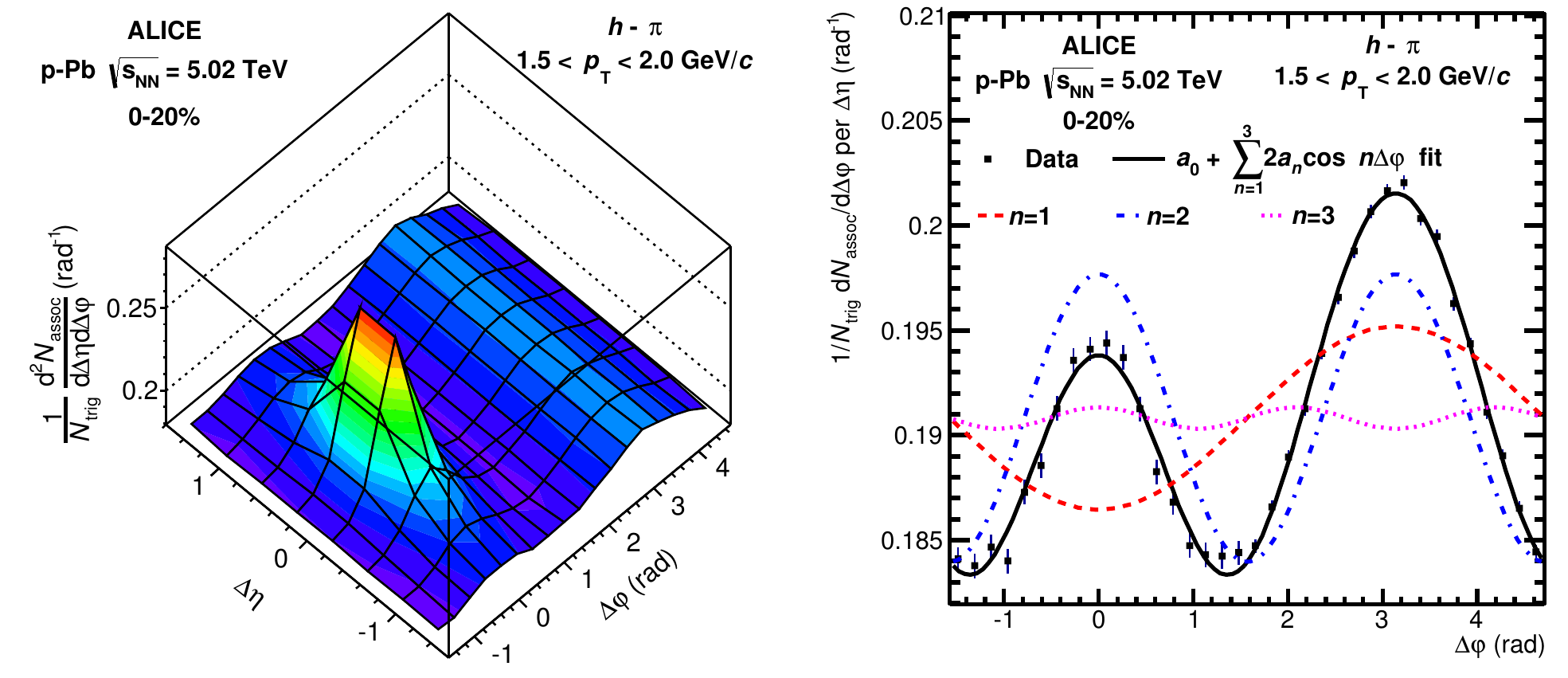}
\caption{\label{fig:pertriggeryields_unsubtracted}
Left panel: associated yield per trigger particle as a function of $\Dphi$ and $\Deta$ for $h-\pi$ correlations with $1.5<\pt<$~\unit[2]{\gevc} in the 0--20\% event class. Right panel: projection of the left panel correlation onto $\Dphi$ averaged over $0.8 < |\Deta| < 1.6$ on the near side and $|\Deta| < 1.6$ on the away side. The fit using Eq.~\ref{fitfunction1} and its individual components are superimposed. The figure contains only statistical uncertainty. Systematic uncertainties are mostly correlated and are less than 5\%.
}
\end{figure}

The left panel of Fig.~\ref{fig:pertriggeryields_unsubtracted} shows the associated yield per
trigger particle for $h-\pi$ correlations with $1.5<\pt<$~\unit[2]{\gevc} in the 0--20\% event
class. On the near side ($|\Dphi| < \pi/2$) a peak originating mostly from jet fragmentation is visible around $\Deta
\approx 0$. In addition, at large $|\Deta|$, the near-side ridge contribution can be observed.
A similar ridge is also present on the away side ($\pi/2 < \Dphi < 3\pi/2$), but it cannot be distinguished from the recoil jet contribution as shown in \cite{alice_pa_ridge}, since both are elongated in $\Deta$.
A similar picture holds for $h-h$, $h-{\rm K}$ and $h-{\rm p}$ correlations. The per-trigger
yield is projected onto $\Dphi$ (right panel of Fig.~\ref{fig:pertriggeryields_unsubtracted})
excluding the near-side peak
by averaging over $0.8 < |\Deta| < 1.6$ on the near side, while on the away
side the average over the full range is used. This $\eta$-gap reduces the jet contribution on the near side, while the away-side jet contribution is still present.

Before further reducing the jet
contribution as in \Ref{alice_pa_ridge}, it is interesting to study the Fourier coefficients extracted from the $\Dphi$ projections. For their determination, these projections are fit with \Eq{fitfunction1}.
This fit describes the data well and is shown in the right panel of Fig.~\ref{fig:pertriggeryields_unsubtracted}. The $\chi^{2}/{\rm ndf}$ is about 0.5--1.5 for all the particle species and $\pt$ intervals.
The first harmonic is found to be negative and contains a contribution from the away-side jet. The second harmonic
has a similar magnitude as the first while the third is much smaller. Including harmonics higher than the third does not change the fit results or the $\chi^2$ significantly. In the following, the
third harmonic is not discussed because the extracted $v_3$ for kaons and protons have large uncertainties such that firm conclusions cannot be drawn.

\begin{figure}[ht!f]
\centering
\includegraphics[width=\textwidth,clip=true,trim=0 0 30 30]{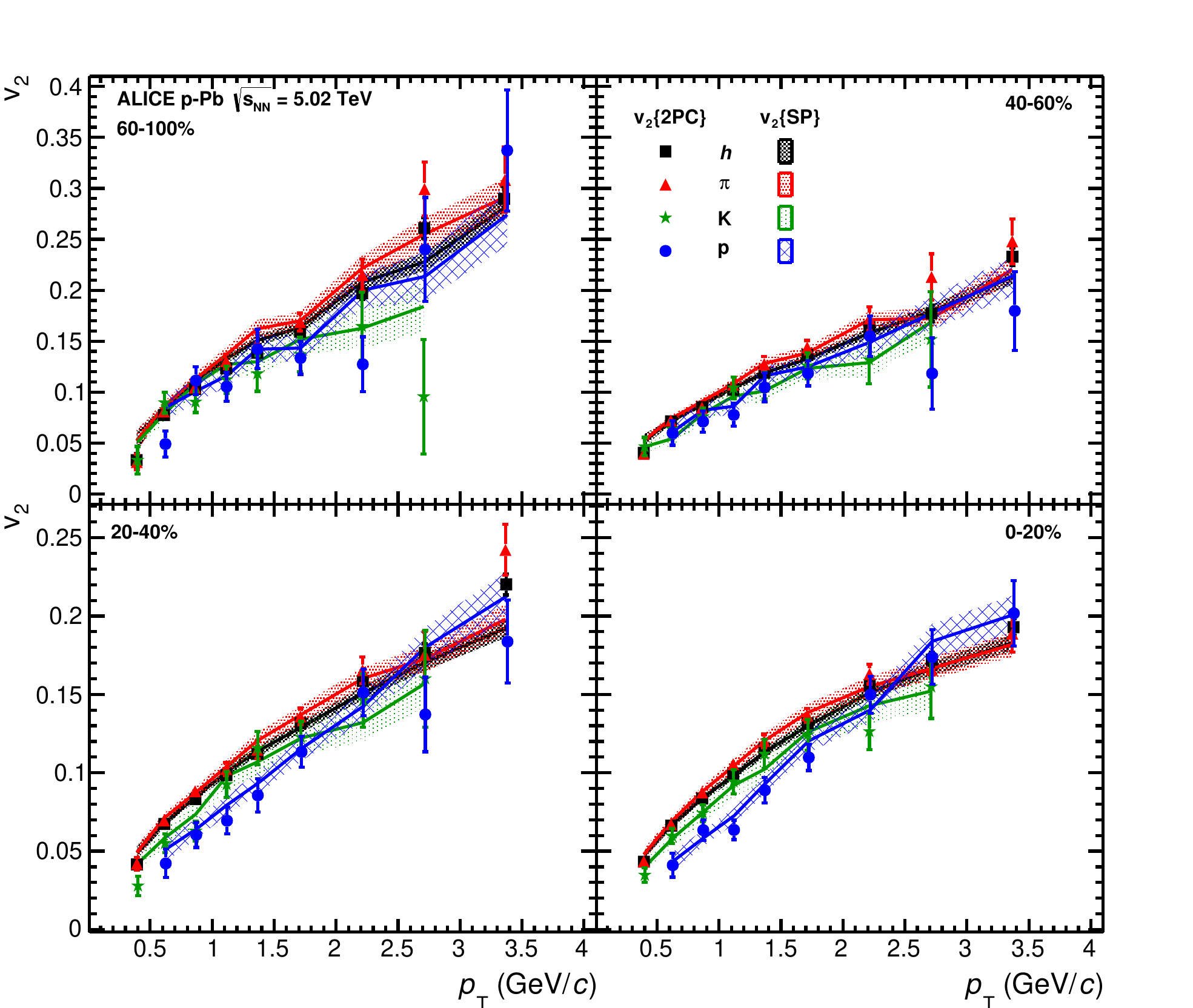}
\caption{\label{fig:v2_unsubtracted}
The Fourier coefficient $v_2$ for all charged particles, pions, kaons and protons as a function of $\pt$ is shown for the different multiplicity classes extracted for $v_2\{{\rm 2PC}\}$ (symbols) and $v_2\{{\rm SP}\}$ (shaded bands, with a line connecting the central values). The data is plotted at the average-$\pt$ for each considered $\pt$ interval and particle species under study. Error bars and widths of the bands show statistical uncertainties and systematic uncertainties, essentially uncorrelated in $\pt$, added in quadrature.
}
\end{figure}

Figure~\ref{fig:v2_unsubtracted} shows the $v_2$ coefficients for $h$,
$\pi$, K and p as a function of $\pt$ for the different multiplicity classes extracted using two-particle correlations ($v_2\{{\rm 2PC}\}$) and the scalar-product method ($v_2\{{\rm SP}\}$). Both methods are generally in good agreement independent of the multiplicity class and the particle species. Large deviations (up to about 30\%) are only observed below \unit[0.5]{\gevc} and for the two lowest-multiplicity classes. At higher $\pt$ and in the higher-multiplicity classes, the agreement between the two methods is better than 10\%.

In the 60--100\% multiplicity class, the $v_2$ coefficients of all the studied particle species are similar and increase as a function of $\pt$. There is a trend of $v_2^{\rm p}$\{SP\} being slightly lower than $v_2^\pi$\{SP\} below \unit[2.5]{\gevc} albeit within the uncertainties. This behaviour in low-multiplicity \pPb\ collisions is qualitatively similar
to that in minimum-bias pp collisions at $\s = \unit[7]{TeV}$ (not shown) where the jet contribution dominates.
Towards higher multiplicities, a different picture emerges. In particular, in the 0--20\% and the 20--40\% multiplicity classes, the particle species are
better separated, with $v_2^{\rm p} < v_2^\pi$ up to about \unit[2]{\gevc}. There is a hint of $v_2^{\rm K} < v_2^\pi$ below \unit[1]{\gevc}. At higher $\pt$, $v^{\rm p}_2\{{\rm SP}\}$ is slightly larger (about $1\sigma$ in the 0--20\% event class) than that of pions, while in the case of $v^{\rm p}_2\{{\rm 2PC}\}$ the uncertainties are too large for a conclusion.

\begin{figure}[ht!f]
\centering
\includegraphics[width=\textwidth]{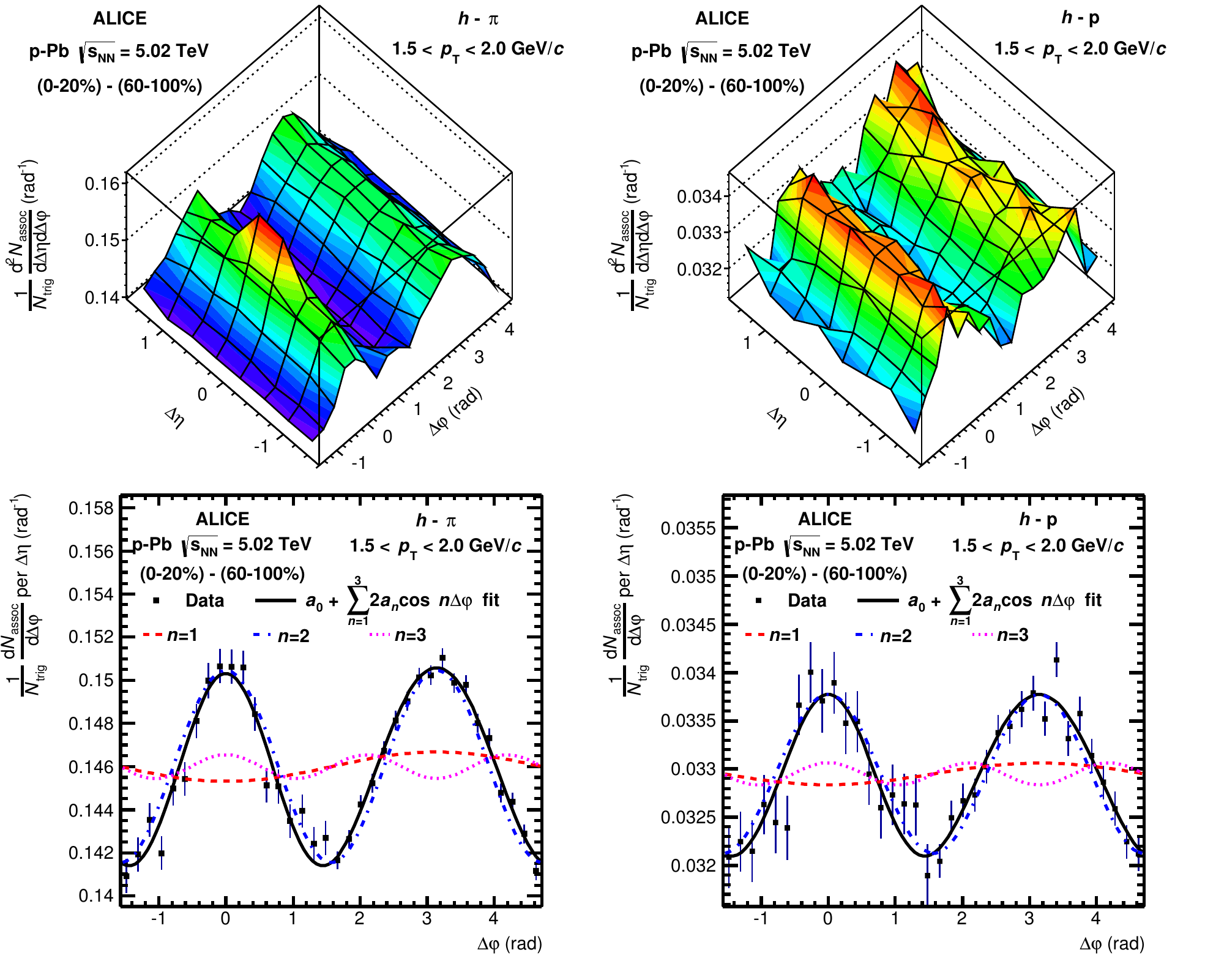}
\caption{\label{fig:pertriggeryields_subtracted}
Top panels: associated yield per trigger particle as a function of $\Dphi$ and $\Deta$ for $h-\pi$ correlations (left) and $h-{\rm p}$ correlations (right) for $1.5<\pt<$~\unit[2]{\gevc} for the 0--20\% event class where the corresponding correlation from the 60--100\% event class has been subtracted.
Bottom panels: projection of the top panel correlations to $\Dphi$ averaged over $0.8 < |\Deta| < 1.6$ on the near side and $|\Deta| < 1.6$ on the away side. The figure contains only statistical uncertainty. Systematic uncertainties are mostly correlated and are less than 5\%.
}
\end{figure}

To further investigate this interesting evolution with multiplicity, the subtraction
method introduced in \Ref{alice_pa_ridge}, which removes a
significant fraction of the correlation due to jets, is applied. The per-trigger yield of the 60--100\%
event class is subtracted from that in the 0--20\% event class. In the upper panels of
Fig.~\ref{fig:pertriggeryields_subtracted} the resulting $h-\pi$ and $h-{\rm p}$ correlation for
$1.5<\pt<$~\unit[2]{\gevc} are shown. In all considered $\pt$-intervals and for all associated particles ($h$, $\pi$, K
and p) a double-ridge structure is observed with a near-side ridge centred at $\Dphi = 0$ and an
away-side ridge centred at $\Dphi = \pi$. Both are independent of $\Deta$ within the studied range
of $|\Deta| < 1.6$, apart from an additional excess which is visible around $\Dphi = \Deta = 0$. This excess is more pronounced for pions than for kaons (not shown) and protons.
This effect is a residue of the jet peak originating in an incomplete subtraction, possibly due to a bias of the event selection on the jet fragmentation. Pions, which are most abundant, are most sensitive to this effect. This residual peak on the near side is excluded by the selection $|\Deta| >
0.8$ when the subtracted correlation is projected onto $\Dphi$. On the away side the full
$\Deta$-range is projected and a residual jet contribution cannot be excluded.
The effect of this residual jet contribution on the measurement is assessed as in \cite{alice_pa_ridge} by:
a) changing the range for the near-side exclusion region from $|\Deta| >
0.8$ to 0.5 and 1.2;
b) subtracting the near-side excess distribution above the ridge also from the away side by reflecting it at $\Dphi = \pi/2$ and scaling it
according to the $\pt$-dependent difference of near-side and away-side jet yields (this difference arises due to the kinematic constraints and the detector acceptance and is evaluated using the lowest-multiplicity class);
and c) scaling the per-trigger yield in the 60--100\% event class such that no near-side peak remains.
The differences in the extracted quantities are included in the systematic uncertainties (3--10\% depending on $\pt$ and particle species).

The lower panels of Fig.~\ref{fig:pertriggeryields_subtracted} show
the $\Dphi$-projections averaged in the same $\Deta$ regions as used for \Fig{fig:pertriggeryields_unsubtracted}.
As before, the Fourier coefficients are extracted from these projections by a fit with Eq.~\ref{fitfunction1}.
These fits are also shown in the lower panels of Fig.~\ref{fig:pertriggeryields_subtracted}. Their $\chi^{2}/{\rm ndf}$ is about 0.6--1.3 for all particle species in the \pt\ range considered, showing that the data is well described by these three Fourier coefficients.
Compared to the case without subtraction, the first Fourier coefficient is up to 10 times smaller, as expected as a consequence of
the significant reduction of the jet component, achieved with the subtraction procedure. The $v_2$ coefficients reduce as well, but only by about 20--40\%. A larger change is seen for protons at low $\pt$.

As already noted in \Ref{alice_pa_ridge} for unidentified particles, no significant near-side ridge is observed in the 60--100\% multiplicity class and it is assumed that the double-ridge structure is not present in this event class.
In the subtraction, along with the jet component, a part of the combinatorial baseline is  removed. This has to be taken into account
when the coefficients $V_{n\Delta}$, which are relative quantities, are extracted.
The $V_{n\Delta}$ coefficients can be extracted from the fit parameters $a_n$ with
$V_{n\Delta}\{{\rm 2PC,sub}\} = a_n / (a_0 + b)$ where the baseline $b$ is the combinatorial baseline of the lower-multiplicity class which has been subtracted ($b$ is determined on the near side within $1.2 < |\Deta| < 1.6$).
From the $V_{n\Delta}^{h-i}$ extracted for the different particle-species combinations, $v_n^i$ is obtained with \Eq{vn}.

\begin{figure}[ht!f]
\centering
\includegraphics[width=0.9\textwidth]{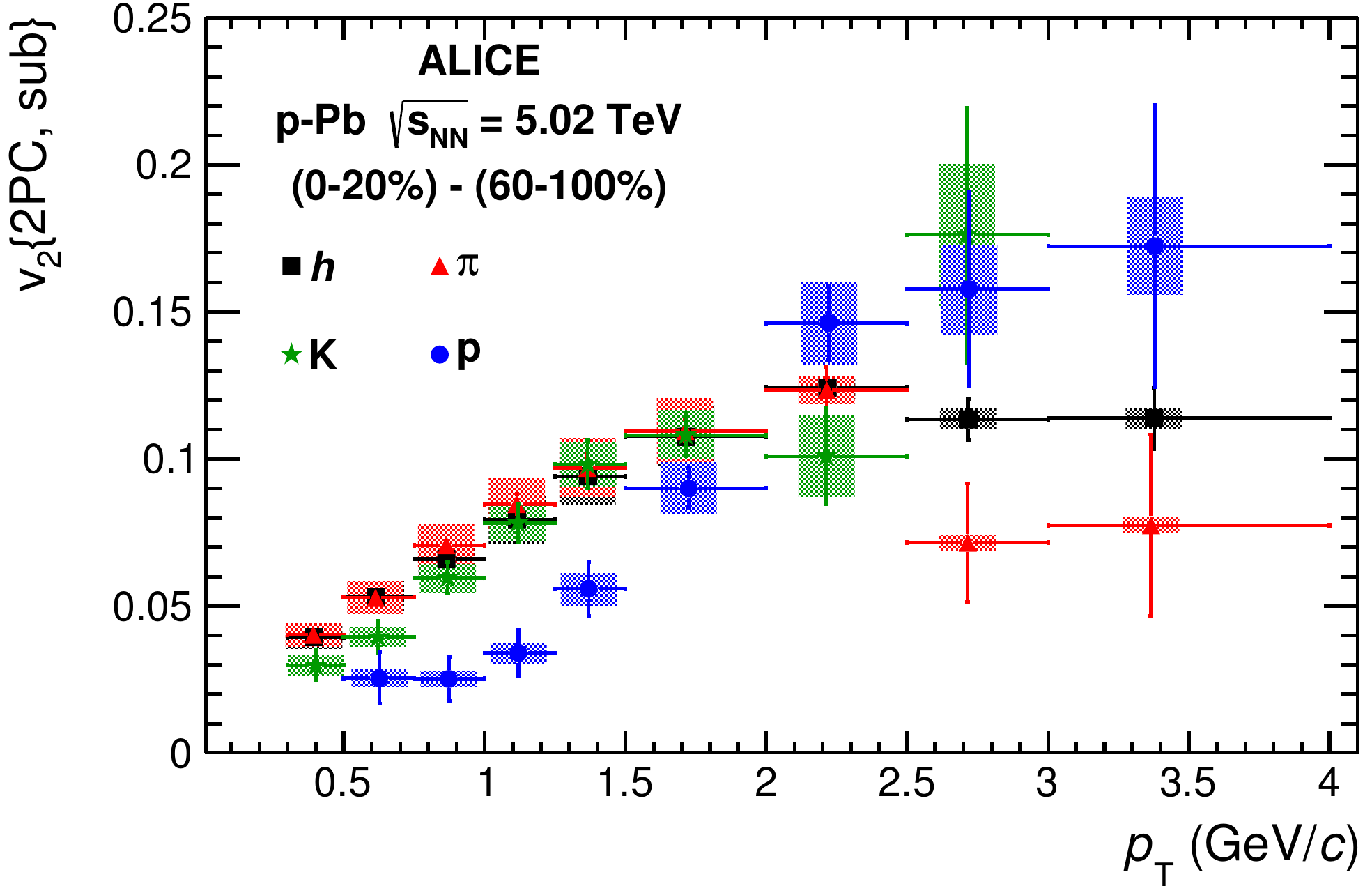}
\caption{\label{fig:v2_subtracted}
The Fourier coefficient $v_2\{{\rm 2PC,sub}\}$ for hadrons (black squares), pions (red triangles), kaons (green
stars) and protons (blue circles) as a function of $\pt$ from the correlation in the 0--20\%
multiplicity class after subtraction of the correlation from the 60--100\% multiplicity class. The data is plotted at the average-$\pt$ for each considered $\pt$ interval and
particle species under study. Error bars show statistical
uncertainties while shaded areas denote systematic uncertainties.
}
\end{figure}

Figure~\ref{fig:v2_subtracted} shows the extracted $v_2\{{\rm 2PC,sub}\}$ coefficients for $h$, $\pi$, K and p
as a function of $\pt$. The coefficient $v_2^{\rm p}$ is significantly lower than
$v_2^\pi$ for $0.5 < \pt <$~\unit[1.5]{\gevc}, and larger than $v_2^\pi$ for $\pt >$~\unit[2.5]{\gevc}. The crossing occurs at $\pt \approx$ \unit[2]{\gevc}. The coefficient $v_2^{\rm K}$ is consistent
with $v_2^\pi$ above \unit[1]{\gevc}; below \unit[1]{\gevc} there is a hint that $v_2^{\rm K}$ is lower than $v_2^\pi$.
The observed behaviour is rather different from that in the 60--100\% multiplicity class (see the top left panel of Fig.~\ref{fig:v2_unsubtracted}) or in pp collisions, which are assumed to be mainly jet-dominated.
The observation of a clear mass ordering between the $v_2$ of pions and protons including their crossing and the hint for a difference between the $v_2$ of pions and kaons is rather intriguing. The mass ordering and crossing is qualitatively similar to observations in nucleus--nucleus collisions \cite{Adler:2003kt,Adams:2004bi,Abelev:2007qg,Heinz:2011kt,Abelev:2012di}. Furthermore, in \Aa\ collisions a mass ordering at low transverse momenta can be described by hydrodynamic model calculations \cite{Huovinen:2001cy, Shen:2011eg}.

The reported results are consistent under a range of variations to the analysis procedure. Changing the multiplicity class for the subtraction to 70--100\% leads to large statistical fluctuations, in particular for protons and kaons. For hadrons and pions the $v_2$ coefficients change by about 8\% below \unit[0.5]{\gevc} and less than 4\% for larger $\pt$.
Repeating the analysis using the 20--40\% event class and subtracting the 60--100\% event class, results in qualitatively similar observations. On average the $v_2$ values are 15--25\% lower and the statistical uncertainties are about a factor 2 larger than in the 0--20\% case. For the 40--60\% event class, the statistical uncertainties are too large to draw a conclusion.

The analysis was repeated using the energy deposited in the \ZNA\ instead of the \VZEROA\ to define the event classes. The extracted $v_2$ values are consistently lower by about 12\% due to the different event sample selected in this way. However, the presented conclusions, in particular the observed difference of $v_2^{\rm p}$ and $v_2^\pi$ compared between jet-dominated
correlations (60--100\% event class) and double-ridge dominated correlations (0--20\% event class after subtraction), are unchanged.

\section{Summary}
\label{sec:summary}
Two-particle angular correlations of charged particles with pions, kaons and protons have been measured in \pPb\ collisions at $\snn=$~\unit[5.02]{TeV} and expressed as associated yields per trigger particle. The Fourier coefficient $v_2$ was extracted from these correlations and studied as a function of $\pt$ and event multiplicity. In low-multiplicity collisions the $\pt$ and species dependence of $v_2$ resembles that observed in pp collisions at similar energy where correlations from jets dominate the measurement. In high-multiplicity \pPb\ collisions a different picture emerges, where $v_2^{\rm p} < v_2^\pi$ is found up to about \unit[2]{\gevc}.
At \unit[3--4]{\gevc}, $v_2^{\rm p}$ is slightly larger than $v_2^\pi$, albeit with low significance.

The per-trigger yield measured in low-multiplicity collisions is subtracted from that measured in high-multiplicity collisions, revealing that the double-ridge structure previously observed in correlations of unidentified particles, is present also in correlations with $\pi$, K and p. The Fourier coefficient $v_2$ of these double-ridge structures exhibits a dependence on $\pt$ that is reminiscent of the one observed in collectivity-dominated Pb--Pb collisions at the LHC: $v_2^{\rm p}$ is significantly smaller than $v_2^\pi$ and $v_2^{\rm K}$ at low $\pt$ while the opposite is observed at \unit[2.5--4]{\gevc}; the crossing takes place at about \unit[2]{\gevc}.

These observations and their qualitative similarity to measurements in \Aa\ collisions are rather intriguing. Their theoretical interpretation is promising to give further insight into the unexpected phenomena observed in \pPb\ collisions at the LHC.

%% file: acknowledgements_march2013.tex
The ALICE collaboration would like to thank all its engineers and technicians for their invaluable contributions to the construction of the experiment and the CERN accelerator teams for the outstanding performance of the LHC complex.
\\
The ALICE collaboration acknowledges the following funding agencies for their support in building and
running the ALICE detector:
 \\
State Committee of Science,  World Federation of Scientists (WFS)
and Swiss Fonds Kidagan, Armenia,
 \\
Conselho Nacional de Desenvolvimento Cient\'{\i}fico e Tecnol\'{o}gico (CNPq), Financiadora de Estudos e Projetos (FINEP),
Funda\c{c}\~{a}o de Amparo \`{a} Pesquisa do Estado de S\~{a}o Paulo (FAPESP);
 \\
National Natural Science Foundation of China (NSFC), the Chinese Ministry of Education (CMOE)
and the Ministry of Science and Technology of China (MSTC);
 \\
Ministry of Education and Youth of the Czech Republic;
 \\
Danish Natural Science Research Council, the Carlsberg Foundation and the Danish National Research Foundation;
 \\
The European Research Council under the European Community's Seventh Framework Programme;
 \\
Helsinki Institute of Physics and the Academy of Finland;
 \\
French CNRS-IN2P3, the `Region Pays de Loire', `Region Alsace', `Region Auvergne' and CEA, France;
 \\
German BMBF and the Helmholtz Association;
\\
General Secretariat for Research and Technology, Ministry of
Development, Greece;
\\
Hungarian OTKA and National Office for Research and Technology (NKTH);
 \\
Department of Atomic Energy and Department of Science and Technology of the Government of India;
 \\
Istituto Nazionale di Fisica Nucleare (INFN) and Centro Fermi -
Museo Storico della Fisica e Centro Studi e Ricerche "Enrico
Fermi", Italy;
 \\
MEXT Grant-in-Aid for Specially Promoted Research, Ja\-pan;
 \\
Joint Institute for Nuclear Research, Dubna;
 \\
National Research Foundation of Korea (NRF);
 \\
CONACYT, DGAPA, M\'{e}xico, ALFA-EC and the EPLANET Program
(European Particle Physics Latin American Network)
 \\
Stichting voor Fundamenteel Onderzoek der Materie (FOM) and the Nederlandse Organisatie voor Wetenschappelijk Onderzoek (NWO), Netherlands;
 \\
Research Council of Norway (NFR);
 \\
Polish Ministry of Science and Higher Education;
 \\
National Authority for Scientific Research - NASR (Autoritatea Na\c{t}ional\u{a} pentru Cercetare \c{S}tiin\c{t}ific\u{a} - ANCS);
 \\
Ministry of Education and Science of Russian Federation, Russian
Academy of Sciences, Russian Federal Agency of Atomic Energy,
Russian Federal Agency for Science and Innovations and The Russian
Foundation for Basic Research;
 \\
Ministry of Education of Slovakia;
 \\
Department of Science and Technology, South Africa;
 \\
CIEMAT, EELA, Ministerio de Econom\'{i}a y Competitividad (MINECO) of Spain, Xunta de Galicia (Conseller\'{\i}a de Educaci\'{o}n),
CEA\-DEN, Cubaenerg\'{\i}a, Cuba, and IAEA (International Atomic Energy Agency);
 \\
Swedish Research Council (VR) and Knut $\&$ Alice Wallenberg
Foundation (KAW);
 \\
Ukraine Ministry of Education and Science;
 \\
United Kingdom Science and Technology Facilities Council (STFC);
 \\
The United States Department of Energy, the United States National
Science Foundation, the State of Texas, and the State of Ohio.

%% file: refpreprint.tex
\providecommand{\href}[2]{#2}\begingroup\raggedright\endgroup

%% file: Alice_Authorlist_2013-Jun-25.tex


\begingroup
\small
\begin{flushleft}
B.~Abelev\Irefn{org69}\And
J.~Adam\Irefn{org36}\And
D.~Adamov\'{a}\Irefn{org77}\And
A.M.~Adare\Irefn{org125}\And
M.M.~Aggarwal\Irefn{org81}\And
G.~Aglieri~Rinella\Irefn{org33}\And
M.~Agnello\Irefn{org104}\textsuperscript{,}\Irefn{org87}\And
A.G.~Agocs\Irefn{org124}\And
A.~Agostinelli\Irefn{org25}\And
Z.~Ahammed\Irefn{org120}
% ...
\renewcommand\labelenumi{\textsuperscript{\theenumi}~}

\section*{Affiliation notes}
\renewcommand\theenumi{\roman{enumi}}
\begin{Authlist}
\item \Adef{0}Deceased
\item \Adef{idp3706720}{Also at: M.V.Lomonosov Moscow State University, D.V.Skobeltsyn Institute of Nuclear Physics, Moscow, Russia}
\item \Adef{idp3950960}{Also at: University of Belgrade, Faculty of Physics and "Vin\v{c}a" Institute of Nuclear Sciences, Belgrade, Serbia}
\item \Adef{idp4245776}{Permanent address: Konkuk University, Seoul, Korea}
\item \Adef{idp4777920}{Also at: Institute of Theoretical Physics, University of Wroclaw, Wroclaw, Poland}
\end{Authlist}

\section*{Collaboration Institutes}
\renewcommand\theenumi{\arabic{enumi}~}
\begin{Authlist}

\item \Idef{org1}A. I. Alikhanyan National Science Laboratory (Yerevan Physics Institute) Foundation, Yerevan, Armenia
\item \Idef{org2}Benem\'{e}rita Universidad Aut\'{o}noma de Puebla, Puebla, Mexico
\item \Idef{org3}Bogolyubov Institute for Theoretical Physics, Kiev, Ukraine
\item \Idef{org4}Budker Institute for Nuclear Physics, Novosibirsk, Russia
\item \Idef{org5}California Polytechnic State University, San Luis Obispo, California, United States
\item \Idef{org6}Central China Normal University, Wuhan, China
\item \Idef{org7}Centre de Calcul de l'IN2P3, Villeurbanne, France 
\item \Idef{org8}Centro de Aplicaciones Tecnol\'{o}gicas y Desarrollo Nuclear (CEADEN), Havana, Cuba
\item \Idef{org9}Centro de Investigaciones Energ\'{e}ticas Medioambientales y Tecnol\'{o}gicas (CIEMAT), Madrid, Spain
\item \Idef{org10}Centro de Investigaci\'{o}n y de Estudios Avanzados (CINVESTAV), Mexico City and M\'{e}rida, Mexico
\item \Idef{org11}Centro Fermi - Museo Storico della Fisica e Centro Studi e Ricerche ``Enrico Fermi'', Rome, Italy
\item \Idef{org12}Chicago State University, Chicago, United States
\item \Idef{org13}Commissariat \`{a} l'Energie Atomique, IRFU, Saclay, France
\item \Idef{org14}COMSATS Institute of Information Technology (CIIT), Islamabad, Pakistan
\item \Idef{org15}Departamento de F\'{\i}sica de Part\'{\i}culas and IGFAE, Universidad de Santiago de Compostela, Santiago de Compostela, Spain
\item \Idef{org16}Department of Physics Aligarh Muslim University, Aligarh, India
\item \Idef{org17}Department of Physics and Technology, University of Bergen, Bergen, Norway
\item \Idef{org18}Department of Physics, Ohio State University, Columbus, Ohio, United States
\item \Idef{org19}Department of Physics, Sejong University, Seoul, South Korea
\item \Idef{org20}Department of Physics, University of Oslo, Oslo, Norway
\item \Idef{org21}Dipartimento di Fisica dell'Universit\`{a} and Sezione INFN, Cagliari, Italy
\item \Idef{org22}Dipartimento di Fisica dell'Universit\`{a} and Sezione INFN, Trieste, Italy
\item \Idef{org23}Dipartimento di Fisica dell'Universit\`{a} and Sezione INFN, Turin, Italy
\item \Idef{org24}Dipartimento di Fisica dell'Universit\`{a} `La Sapienza` and Sezione INFN, Rome, Italy
\item \Idef{org25}Dipartimento di Fisica e Astronomia dell'Universit\`{a} and Sezione INFN, Bologna, Italy
\item \Idef{org26}Dipartimento di Fisica e Astronomia dell'Universit\`{a} and Sezione INFN, Catania, Italy
\item \Idef{org27}Dipartimento di Fisica e Astronomia dell'Universit\`{a} and Sezione INFN, Padova, Italy
\item \Idef{org28}Dipartimento di Fisica `E.R.~Caianiello' dell'Universit\`{a} and Gruppo Collegato INFN, Salerno, Italy
\item \Idef{org29}Dipartimento di Scienze e Innovazione Tecnologica dell'Universit\`{a} del Piemonte Orientale and Gruppo Collegato INFN, Alessandria, Italy
\item \Idef{org30}Dipartimento Interateneo di Fisica `M.~Merlin' and Sezione INFN, Bari, Italy
\item \Idef{org31}Division of Experimental High Energy Physics, University of Lund, Lund, Sweden
\item \Idef{org32}Eberhard Karls Universit\"{a}t T\"{u}bingen, T\"{u}bingen, Germany
\item \Idef{org33}European Organization for Nuclear Research (CERN), Geneva, Switzerland
\item \Idef{org34}Faculty of Engineering, Bergen University College, Bergen, Norway
\item \Idef{org35}Faculty of Mathematics, Physics and Informatics, Comenius University, Bratislava, Slovakia
\item \Idef{org36}Faculty of Nuclear Sciences and Physical Engineering, Czech Technical University in Prague, Prague, Czech Republic
\item \Idef{org37}Faculty of Science, P.J.~\v{S}af\'{a}rik University, Ko\v{s}ice, Slovakia
\item \Idef{org38}Frankfurt Institute for Advanced Studies, Johann Wolfgang Goethe-Universit\"{a}t Frankfurt, Frankfurt, Germany
\item \Idef{org39}Gangneung-Wonju National University, Gangneung, South Korea
\item \Idef{org40}Helsinki Institute of Physics (HIP), Helsinki, Finland
\item \Idef{org41}Hiroshima University, Hiroshima, Japan
\item \Idef{org42}Indian Institute of Technology Bombay (IIT), Mumbai, India
\item \Idef{org43}Indian Institute of Technology Indore, India (IITI)
\item \Idef{org44}Institut de Physique Nucl\'{e}aire d'Orsay (IPNO), Universit\'{e} Paris-Sud, CNRS-IN2P3, Orsay, France
\item \Idef{org45}Institut f\"{u}r Informatik, Johann Wolfgang Goethe-Universit\"{a}t Frankfurt, Frankfurt, Germany
\item \Idef{org46}Institut f\"{u}r Kernphysik, Johann Wolfgang Goethe-Universit\"{a}t Frankfurt, Frankfurt, Germany
\item \Idef{org47}Institut f\"{u}r Kernphysik, Technische Universit\"{a}t Darmstadt, Darmstadt, Germany
\item \Idef{org48}Institut f\"{u}r Kernphysik, Westf\"{a}lische Wilhelms-Universit\"{a}t M\"{u}nster, M\"{u}nster, Germany
\item \Idef{org49}Institut Pluridisciplinaire Hubert Curien (IPHC), Universit\'{e} de Strasbourg, CNRS-IN2P3, Strasbourg, France
\item \Idef{org50}Institute for High Energy Physics, Protvino, Russia
\item \Idef{org51}Institute for Nuclear Research, Academy of Sciences, Moscow, Russia
\item \Idef{org52}Institute for Subatomic Physics of Utrecht University, Utrecht, Netherlands
\item \Idef{org53}Institute for Theoretical and Experimental Physics, Moscow, Russia
\item \Idef{org54}Institute of Experimental Physics, Slovak Academy of Sciences, Ko\v{s}ice, Slovakia
\item \Idef{org55}Institute of Physics, Academy of Sciences of the Czech Republic, Prague, Czech Republic
\item \Idef{org56}Institute of Physics, Bhubaneswar, India
\item \Idef{org57}Institute of Space Science (ISS), Bucharest, Romania
\item \Idef{org58}Instituto de Ciencias Nucleares, Universidad Nacional Aut\'{o}noma de M\'{e}xico, Mexico City, Mexico
\item \Idef{org59}Instituto de F\'{\i}sica, Universidad Nacional Aut\'{o}noma de M\'{e}xico, Mexico City, Mexico
\item \Idef{org60}iThemba LABS, National Research Foundation, Somerset West, South Africa
\item \Idef{org61}Joint Institute for Nuclear Research (JINR), Dubna, Russia
\item \Idef{org62}Korea Institute of Science and Technology Information, Daejeon, South Korea
\item \Idef{org63}KTO Karatay University, Konya, Turkey
\item \Idef{org64}Laboratoire de Physique Corpusculaire (LPC), Clermont Universit\'{e}, Universit\'{e} Blaise Pascal, CNRS--IN2P3, Clermont-Ferrand, France
\item \Idef{org65}Laboratoire de Physique Subatomique et de Cosmologie (LPSC), Universit\'{e} Joseph Fourier, CNRS-IN2P3, Institut Polytechnique de Grenoble, Grenoble, France
\item \Idef{org66}Laboratori Nazionali di Frascati, INFN, Frascati, Italy
\item \Idef{org67}Laboratori Nazionali di Legnaro, INFN, Legnaro, Italy
\item \Idef{org68}Lawrence Berkeley National Laboratory, Berkeley, California, United States
\item \Idef{org69}Lawrence Livermore National Laboratory, Livermore, California, United States
\item \Idef{org70}Moscow Engineering Physics Institute, Moscow, Russia
\item \Idef{org71}National Centre for Nuclear Studies, Warsaw, Poland
\item \Idef{org72}National Institute for Physics and Nuclear Engineering, Bucharest, Romania
\item \Idef{org73}National Institute of Science Education and Research, Bhubaneswar, India
\item \Idef{org74}Niels Bohr Institute, University of Copenhagen, Copenhagen, Denmark
\item \Idef{org75}Nikhef, National Institute for Subatomic Physics, Amsterdam, Netherlands
\item \Idef{org76}Nuclear Physics Group, STFC Daresbury Laboratory, Daresbury, United Kingdom
\item \Idef{org77}Nuclear Physics Institute, Academy of Sciences of the Czech Republic, \v{R}e\v{z} u Prahy, Czech Republic
\item \Idef{org78}Oak Ridge National Laboratory, Oak Ridge, Tennessee, United States
\item \Idef{org79}Petersburg Nuclear Physics Institute, Gatchina, Russia
\item \Idef{org80}Physics Department, Creighton University, Omaha, Nebraska, United States
\item \Idef{org81}Physics Department, Panjab University, Chandigarh, India
\item \Idef{org82}Physics Department, University of Athens, Athens, Greece
\item \Idef{org83}Physics Department, University of Cape Town, Cape Town, South Africa
\item \Idef{org84}Physics Department, University of Jammu, Jammu, India
\item \Idef{org85}Physics Department, University of Rajasthan, Jaipur, India
\item \Idef{org86}Physikalisches Institut, Ruprecht-Karls-Universit\"{a}t Heidelberg, Heidelberg, Germany
\item \Idef{org87}Politecnico di Torino, Turin, Italy
\item \Idef{org88}Purdue University, West Lafayette, Indiana, United States
\item \Idef{org89}Pusan National University, Pusan, South Korea
\item \Idef{org90}Research Division and ExtreMe Matter Institute EMMI, GSI Helmholtzzentrum f\"ur Schwerionenforschung, Darmstadt, Germany
\item \Idef{org91}Rudjer Bo\v{s}kovi\'{c} Institute, Zagreb, Croatia
\item \Idef{org92}Russian Federal Nuclear Center (VNIIEF), Sarov, Russia
\item \Idef{org93}Russian Research Centre Kurchatov Institute, Moscow, Russia
\item \Idef{org94}Saha Institute of Nuclear Physics, Kolkata, India
\item \Idef{org95}School of Physics and Astronomy, University of Birmingham, Birmingham, United Kingdom
\item \Idef{org96}Secci\'{o}n F\'{\i}sica, Departamento de Ciencias, Pontificia Universidad Cat\'{o}lica del Per\'{u}, Lima, Peru
\item \Idef{org97}Sezione INFN, Bari, Italy
\item \Idef{org98}Sezione INFN, Bologna, Italy
\item \Idef{org99}Sezione INFN, Cagliari, Italy
\item \Idef{org100}Sezione INFN, Catania, Italy
\item \Idef{org101}Sezione INFN, Padova, Italy
\item \Idef{org102}Sezione INFN, Rome, Italy
\item \Idef{org103}Sezione INFN, Trieste, Italy
\item \Idef{org104}Sezione INFN, Turin, Italy
\item \Idef{org105}SUBATECH, Ecole des Mines de Nantes, Universit\'{e} de Nantes, CNRS-IN2P3, Nantes, France
\item \Idef{org106}Suranaree University of Technology, Nakhon Ratchasima, Thailand
\item \Idef{org107}Technical University of Split FESB, Split, Croatia
\item \Idef{org108}The Henryk Niewodniczanski Institute of Nuclear Physics, Polish Academy of Sciences, Cracow, Poland
\item \Idef{org109}The University of Texas at Austin, Physics Department, Austin, TX, United States
\item \Idef{org110}Universidad Aut\'{o}noma de Sinaloa, Culiac\'{a}n, Mexico
\item \Idef{org111}Universidade de S\~{a}o Paulo (USP), S\~{a}o Paulo, Brazil
\item \Idef{org112}Universidade Estadual de Campinas (UNICAMP), Campinas, Brazil
\item \Idef{org113}University of Houston, Houston, Texas, United States
\item \Idef{org114}University of Jyv\"{a}skyl\"{a}, Jyv\"{a}skyl\"{a}, Finland
\item \Idef{org115}University of Tennessee, Knoxville, Tennessee, United States
\item \Idef{org116}University of Tokyo, Tokyo, Japan
\item \Idef{org117}University of Tsukuba, Tsukuba, Japan
\item \Idef{org118}Universit\'{e} de Lyon, Universit\'{e} Lyon 1, CNRS/IN2P3, IPN-Lyon, Villeurbanne, France
\item \Idef{org119}V.~Fock Institute for Physics, St. Petersburg State University, St. Petersburg, Russia
\item \Idef{org120}Variable Energy Cyclotron Centre, Kolkata, India
\item \Idef{org121}Vestfold University College, Tonsberg, Norway
\item \Idef{org122}Warsaw University of Technology, Warsaw, Poland
\item \Idef{org123}Wayne State University, Detroit, Michigan, United States
\item \Idef{org124}Wigner Research Centre for Physics, Hungarian Academy of Sciences, Budapest, Hungary
\item \Idef{org125}Yale University, New Haven, Connecticut, United States
\item \Idef{org126}Yonsei University, Seoul, South Korea
\item \Idef{org127}Zentrum f\"{u}r Technologietransfer und Telekommunikation (ZTT), Fachhochschule Worms, Worms, Germany
\end{Authlist}
\endgroup